\documentclass[aps,prl,twocolumn,tightenlines,superscriptaddress]{revtex4} 
\usepackage{epsfig}
\usepackage{graphicx}
\usepackage{psfrag}
\usepackage{amsmath,amssymb}
\usepackage{colordvi}
\usepackage{amsfonts}
\usepackage{enumerate}
\usepackage{slashed}

\begin{document}

\title{Comment on ``Spin and Orbital Angular Momentum in Gauge Theories" by X. S. Chen et. al. (PRL100, 232002 (2008))}
\author{Xiangdong Ji}
\affiliation{Department of Physics, University of Maryland, College
Park, Maryland 20742, USA} \affiliation{Center for High-Energy
Physics and Institute of Theoretical Physics, \\ Peking University
Beijing, 100080, P. R. China}

\date{\today}
\vspace{0.5in}
\begin{abstract}
The individual parts of the total angular momentum operator
in interacting theories cannot satisfy the canonical angular momentum
commutation rule, including those proposed in the above paper.
Furthermore, the operators in the new proposal 
a) are non-local in general gauge, b) do not have proper Lorentz transformation
properties, and c) do not have any known physical measurements.
\end{abstract}

\maketitle

In a recent article \cite{Chen:2008ag}, Chen et al. claim to have solved
a ``long-standing gauge-invariance problem of the nucleon spin structure" and
``this was previously thought to be an impossible task." 
In this comment, I would like to point out that the
solution proposed  in \cite{Chen:2008ag} is less satisfactory than the
authors have thought.

According to \cite{Chen:2008ag}, the {\it long-standing problem} is to find out 
the {\it appropriate} operators for the spin and orbital angular momentum (OAM) of the
quark and gluon fields. A careful reading of the paper finds two specific 
conditions: 1) gauge invariance and 2) individual parts obeying the angular momentum
algebra, $[J_i, J_j] = i\hbar \epsilon_{ijk} J_k$.  The first condition is 
of course a must,  but the second requirement is an ``impossible task" in interacting 
quantum field theories (QFTs), and the operators proposed in 
Ref. \cite{Chen:2008ag} certainly do not satisfy this, contrary to their claim. 
The argument is relatively simple:

The total angular momentum operators $\vec{J}$ of any Lorentz-invariant
field theory obey the angular momentum algebra. However, when breaking them into sums
of parts,
  $ \vec{J} = \sum_i \vec{J}_i(\mu), $
$\vec{J}_i(\mu)$ are not conserved and hence depend on the
renormalization scale $\mu$ when non-trivial interactions are present.
Therefore, even when $\vec{J}_i(\mu)$ satisfy the angular momentum relations
at $\mu_1$, they ought be violated at scale $\mu_2$ as both sides of
the equation necessarily have different $\mu$ dependence.
Since there is no one scale more superior than other,
the angular momentum relations for the individual parts
are generally untrue. This is the case even for
$\psi^\dagger\frac{\vec{\Sigma}}{2}\psi$ because the singlet axial current
is not conserved and has an anomalous dimension due to the well-known
Adler-Bell-Jackiw anomaly~\cite{Peskin:1995ev}.

Now turn to the more delicate issue of gauge invariance. Chen et al. proposed
a new gauge-invariant representation of the QED and QCD angular momenta~\cite{Chen:2008ag}.
However, what they regard as ``gauge-invariant" operators are not the usual
textbook type, and the new notion of gauge invariance clashes with locality and Lorentz symmetry, 
and ultimately has limited physical significance and value. 

First, the so-called ``physical" part of
the gauge potential defined in Ref.~\cite{Chen:2008ag} has the following manifest non-local solution,
\begin{eqnarray}
   \vec{A}_{\rm phys} &=& \vec{A} - \frac{\vec{\nabla}}{\nabla^2} \vec{\nabla} \cdot \vec{A}
\end{eqnarray}
Thus the quark orbital angular momentum (OAM), the gluon spin, and gluon OAM as 
they defined, although formally gauge-invariant, are all non-local operators of the gauge potential. 
In local QFT's, the gauge symmetry impose important constraints, however, if locality is ignored, many
quantities can be made trivially gauge-invariant by simply adding a gauge link to infinity. 


A more serious problem is Lorentz symmetry: If a gauge potential $\vec{A}_{\rm phys}$ 
satisfies the ``physical condition" in one frame $\vec{\nabla} \cdot \vec{A}_{\rm phys}=0$, the 
transformed potential no longer satisfies the condition $\vec{\nabla}' \cdot \vec{A}'_{\rm phys}\ne 0$ 
in a different frame. Thus, observables in terms of the ``physical part" of $\vec{A}$ in 
one frame will in general contain non-physical contribution as seen by different observers! 
Moreover, these non-local observables do not in general transform as Lorentz 
scalar, vector, or tensor. To give a specific example,
the matrix element of the gluon spin operator $\int d^3 x \vec{E} \times \vec{A}_{\rm phys}$
in \cite{Chen:2008ag} depends on the speed of the parent hadron, even when
measured in helicity eigenstates. There is no transformation rule relating the
spin matrix elements of different frames, making the notion of gluon spin
contribution non-intrinsic and less physical.

Gauge theories in the textbooks construct physical observables as gauge-invariant 
{\it local operators}. When solving a specific problem, such as hydrogen atom, one can choose a particular 
reference of frame and gauge, the wave functions (electron's or
photon multipoles) are obviously gauge-dependent, but the true physical observables calculated from them
are not and transform properly under Lorentz transformation. An example for
QCD angular momentum is what has been considered in Ref.~\cite{Ji:1996ek}. Ultimately, the physics is 
learnt only when there are real experimental measurements so that one can compare theoretical 
calculations with data. If operators constructed are fundamentally unsound, it is 
doubtful that any experiment can ever be found to measure their matrix elements.

I thank T. Cohen for an interesting discussion. This work was partially supported by the U. S. Department of Energy via grant
DE-FG02-93ER-40762.

\end{document}